# Density_Enhancement_Streams_in_The_Solar_Wind

By F.S. Mozer, O. Agapitov, S.D. Bale, R. Livi, O. Romeo, K. Sauer, I.Y. Vasko and J. Verniero

This letter describes a new phenomenon on the Parker Solar Probe of recurring plasma density enhancements that have $\Delta n/n$ ~10% and that occur at a repetition rate of ~5 Hz. They were observed sporadically for about five hours between 14 and 15 solar radii on Parker Solar Probe orbit 12 and they were also seen in the same radial range on both the inbound and outbound orbits 11. Their apparently steady-state existence suggests that their pressure gradient was balanced by the electric field. The EX electric field component produced from this requirement is in good agreement with that measured. This provides strong evidence for the measurement accuracy of the density fluctuations and the X- and Y-components of the electric field (the Z-component was not measured). The electrostatic density waves were accompanied by an electromagnetic low frequency wave which occurred with the electrostatic harmonics. The amplitudes of these electrostatic and electromagnetic waves at ≥ 1 Hz were greater than the amplitude of the Alfvenic turbulence in their vicinity so they can be important for the heating, scattering, and acceleration of the plasma. The existence of this pair of waves is consistent with the observed plasma distributions and is explained by a magneto-acoustic wave theory that produces a low frequency electromagnetic wave and electrostatic harmonics.

**I Introduction**

Low frequency turbulence is thought to energize the solar wind plasma through a cascade process that is described by the power spectra of the fields. The magnetic field has generally been the parameter utilized in such studies [Bowen et al, 2020; Chen et al, 2010, 2016, 2020, 2021]. In a few cases, the electric field and/or the plasma density spectra have been shown to have decreasing power in the kinetic range as compared to that in the inertial range [Chen et al, 2013; Mozer et al, 2020; Salem et al, 2012]. The purpose of this letter is to describe a new wave mode in which the power in the solar wind plasma density and electric field exceeded that of the local Alfvenic turbulence.

The measurements of interest were made on the Parker Solar Probe, whose X-Y plane, perpendicular to the Sun-satellite line, contains a two-component electric field and spacecraft potential measurement by antennas that are not much larger than the spacecraft [Bale et al, 2016]. By fitting the measured spacecraft potential to the low-rate density measurements obtained from the SWEAP plasma measurements [Kasper et al, 2016; Whittlesey et al, 2019], higher frequency estimates of the plasma density and density fluctuations [Mozer et al, 2022] have been obtained.

**II Data**

Examples of power spectra obtained on Parker Solar Probe orbit 12 during a 15-second-interval when the spacecraft was located about 15 solar radii from the Sun are given in Figure 1a.  The magnetic field spectrum (green) is typical of that often observed, with excess power at ~1 Hz, a decrease in power of three orders of magnitude between 1 and 20 Hz, and a break at 10 Hz, near the ion gyrofrequency (denoted by the vertical dashed line).  In normal turbulence, the electric power decreases by a similar factor over the given frequency range.  However, the electric field (black) and density (red) spectra in the example of Figure 1 differ greatly from this expectation, having peaks near 1, 5, 11, and 16 Hz with powers more than one to three orders-of-magnitude greater than that expected from the magnetic field spectrum.  Because there is power near 1 Hz in all three fields, there must be both an electrostatic and electromagnetic wave at this frequency.  The peaks at higher frequencies must be due to an electrostatic wave whose characteristics are displayed in the waveform plots of panels 1b and 1d as spiky pulses of electric field and density fluctuations, $\Delta n/n$, occurring at a frequency of several Hz. These density waves were observed sporadically for about five hours between 14 and 15 solar radii on this orbit and they were also seen in the same radial range on both the inbound and outbound orbits 11. (Other orbits have not been carefully searched to see if these waves are more wide-spread).  As summarized above, the electric field power in these waves is much greater than that in the Alfvenic turbulence at the same frequencies.

The plasma density is measured by the particle instruments on the Parker Solar Probe at an ~1 Hz rate.  To obtain the plasma density at a higher rate, the spacecraft potential (the average of the four biased antenna voltages), is utilized.  It is measured at a high rate (~500 Hz) and it is approximately proportional to the log of the plasma density [Mozer et al, 2022].  The fit of the low-rate spacecraft potential to the log of the plasma density provides a least-squares equation that is then used with the high-rate spacecraft potential to obtain the high-rate plasma density and its fluctuations.

Figure 2 provides further information on the density waves, in which 2a gives the density fluctuation amplitude, Δn/n~0.1, while 2b and 2c give the two measured electric field components.  The unmeasured Ez component is estimated in panel 2d by combining the two measured components with the assumption that the parallel electric field was zero.  It was produced to show that the major electric field component was in the X-direction.  Because this is an electrostatic wave, its k-vector must also have been largely in the same X-direction.  Because the magnetic field components at this time were (0, -200, -600), the wave propagated nearly perpendicular to the background magnetic field.

The long duration of these density streams suggest that they are stable structures.  According to the Generalized Ohm's law, this stability requires an electric field that balances the pressure gradient.  This pressure gradient electric field, in the X-direction, is computed as shown in Figure 2e, for the observed temperature of 50 eV and an X-component solar wind speed of 200 km/sec. (The correct speed to use is the solar wind speed plus the sound wave speed in the plasma frame.  Because the sound wave speed is about three times smaller than the Alfven speed (because beta~0.1), its speed is about three times smaller than the solar wind speed so it is ignored for this computation.)

This estimate of Ex, given in panel 2e, is in reasonable agreement with the measured electric field in panel 2b.  This agreement offers evidence that the electric fields and the density streams were correctly measured and that the density structures were in pressure balance.

Also shown in Figures 2f and 2g are the four single-ended potentials of the electric field measurement.  Their similarity and normal behavior suggest that the resulting fields and density structures were well measured and not associated with a spacecraft wake or similar non-physical perturbation.  All data in Figure 2, other than that in panels 2f and 2g, were high pass filtered at one Hz.

Variations of the proton and electron fluxes and spectra were uncorrelated with the times of the density streams.  In addition, beta (0.1), the Alfven speed divided by the solar wind speed (~1), the ion temperature (50 eV), the wind speed (400-600 km/sec), and the Debye length (2 m), all had significant fluctuations but none of them correlated with the off-and-on nature of Δn/n.

The electrostatic potential in these waves may be estimated from Ohm's law as

$$e\phi \sim T_e \cdot \Delta n/n \sim 0.1 \cdot T_e \qquad (1)$$

where $\phi$ is the potential in the wave. For the observed electron core temperature of 50 eV, the potential in the wave was ~5 Volts. This potential is sufficiently large that it may be at least partially responsible for heating or accelerating the plasma by their trapping in the wave potential.

The waveforms, being spiky, appear to be steepened. The time for a sound wave to steepen is about

$$\tau \sim 1/[k \cdot c_s \cdot \Delta n/n] \qquad (2)$$

where $\tau$ is the steepening time, k is the initial wave number and cs is the sound speed. Because $k \cdot V_{sw} \sim \omega$, where $\omega$ is the observed angular frequency and $V_{sw}/c_s \sim 3$, the steepening time is the order of one second. This explains why the density waves are steepened to appear as shocklets because their lifetimes are observed to be greater than ~one second.

Figure 3 presents another example of the density streams. In this case, there are combined electrostatic and electromagnetic signals at both 1 and 8 Hz, while there are harmonics of the electrostatic signal to frequencies greater than 50 Hz. It is again noted that the amplitude of the electric field is an order-of-magnitude greater than that expected from the turbulence described by the magnetic field.

Additional properties associated with this wave mode are illustrated in Figure 4, which presents data obtained over the one-hour interval that included the density streams. Figures 4a and 4b give wavelet plots of the electric field and density fluctuations which show that the ~5 Hz power of the density streams and the associated ~5 Hz electric field fluctuations occurred sporadically during the time interval. In addition to the ~5 Hz power in the electric field, there was power at ≤1 Hz, which was due to the electric field of the simultaneously occurring electromagnetic wave. As expected, this power was absent in the density fluctuations of figure 4b. Figure 4c presents the core proton perpendicular temperature divided by the core proton parallel temperature ($T_{perp}/T_{par}$), which varied but was as large as 10 sporadically, but not necessarily at the time of the density streams. This result is seen inside the three pairs of vertical dashed lines that border three of the about eight regions of enhanced fields and density fluctuations. Figure 4d gives the proton and electron plasma betas whose values were around 0.1 with the proton beta being somewhat larger than the electron beta.

To summarize the experimental data:
   a. Streams of spiky, enhanced plasma density occurring at a rate of ~5 Hz and harmonics have been observed along with the electric field required for pressure balance in the plasma.
   b. The spectra of the density streams and the associated electric field were comprised of many frequency harmonics.
   c. These electrostatic waves occurred along with a low frequency electromagnetic wave.
   d. The amplitudes of these electrostatic and electromagnetic waves were greater than the amplitude of the nearby Alfvenic turbulence.
   e. The k-vector of the density stream electric field was perpendicular to the background magnetic field.
   f. The core proton temperature distribution was anisotropic with $T_{perp}/T_{par}$ as large as 10 at times in the vicinity of the density streams.
   g. The proton plasma beta was somewhat larger than the electron plasma beta but both were near 0.1.

**III DISCUSSION**

Previous work on magneto-acoustic (magnetosonic) waves has shown that a low frequency electromagnetic ion cyclotron wave (EMIC) can be accompanied by harmonics that are electrostatic [Zhu, et al, 2019; Gao, et al, 2021]. This suggests that magneto-acoustic waves may be associated with the observed density streams and electromagnetic wave, as discussed in the following. Figure 5 presents results of dispersion theory using two approaches. A first good overview of the electromagnetic wave modes existing in the range of the proton cyclotron frequency is obtained using the Hall-MHD theory, as presented in a recent paper about multi-ion plasmas [Sauer and Dubinin, 2022]. Neglecting damping effects, it has the advantage that the dispersion of the wave modes in question can be determined with relatively little effort by solving polynomials in ω=ω(k). For the sake of simplification, only thermal effects of the (massless) electrons have been taken into account there. Based on the wave modes found in this way, the corresponding solutions of the kinetic dispersion theory (Vlasov approach) have been determined, including the relevant plasma parameters of the observations, as electron and proton temperature and their anisotropy. The Fortran code applied here has been used in past for several other purposes, such as for studies of the excitation of whistler waves [Sauer and Sydora, 2010] and EMIC waves [Sauer and Dubinin 2022]. For completeness, the following instability theory has been carried out for both parallel and perpendicular wave propagation although the case of current interest corresponds to the observed perpendicular wave propagation.

The results of the present investigations can be summarized as follows: For *quasi-parallel* propagation, the phase velocity of the magneto-acoustic wave with the dispersion ω=kV$_s$, where V$_s$=(γ$k$T$_e$/m$_p$)$^{1/2}$ is the ion sound speed, is determined by V$_{ph}$/V$_{Ap}$=(0.5γβ$_e$)$^{1/2}$; γ=3 is the adiabatic exponent. For β$_e$=0.15 (as observed) one thus obtains V$_{ph}$/V$_{Ap}$ ≤0.5. This creates the possibility of coupling the magneto-acoustic mode (MA) with the ion-cyclotron wave (L). This happens in the range of kc/ω$_p$≥1 where both waves have approximately the same phase velocities. As a consequence, mode splitting occurs, as clearly seen within the marked circles in panels a$_1$ and b$_1$ of Figure 5. With regard to the real frequency versus k, the Vlasov theory gives completely identical results (panels d$_1$ and e$_1$. With the selected proton anisotropy of T$_{perp}$/T$_{par}$ =6 (as was observed sporadically), both modes become unstable (panel f$_1$). Due to the strong damping of the MA wave within the coupling region, a higher temperature anisotropy is required for the same growth rate of the ion cyclotron (L) wave than in the absence of the MA wave. The instability disappears when the anisotropy drops to about four.

Results of dispersion theory of both approaches for *quasi-perpendicular* propagation (θ=85$^0$) are shown in panels a$_2$ to f$_2$. In the Hall-MHD theory, the three modes are shown, the right-polarized R-mode with V$_{ph}$ ~V$_{Ap}$, the mixed modes L-MA and the MA-L whose phase velocities at k=0 are given by V$_{ph}$=V$_{Ap}$ cosθ and V$_{ph}$=V$_s$ cosθ, respectively. These at the origin (k=0) interchanged values of the phase velocities are an effect of the occurring coupling of both modes by which the mixed character of both modes arises. According to the Vlasov theory, the L-MA mode is strongly damped with increasing wave number, which causes the modification of the real part of the frequency (see panels d$_2$ and e$_2$ of Figure 5) compared to the behavior of the Hall-MHD theory. With sufficient temperature anisotropy, the kinetic damping can be compensated by the increasing solution, as can be seen in panel f$_2$. For the chosen β$_e$=β$_p$=0.15, the threshold of instability is T$_{perp}$/T$_{par}$~4. It should be pointed out that the L-MA mode of the Hall-MHD theory is identical to the Alfvén kinetic wave in low-beta plasmas, which has been studied e.g., by Schekochihin et al. [2009] and Narita et al. [2020]. Accordingly, its phase velocity can be expressed analytically by

$$V_{ph} = V_{Ap} cos\theta \sqrt{\left[1 + \frac{\beta_e}{2}\left(\frac{kc}{\omega_p}\right)^2 sin^2\theta\right]} \qquad (3)$$

which is in agreement with the dependency of the Hall-MHD theory shown in Figure 5 b$_2$.

Apart from the simultaneous excitation of two wave modes by temperature anisotropy, the ion cyclotron wave at quasi-parallel propagation and the (modified) magneto-acoustic wave at quasi-perpendicular propagation, the dispersion analysis revealed another interesting aspect. It deals with the consequences of the coupling of two wave modes in the ω-k range with nearly the same phase velocity. The associated effects have already been investigated for a number of other wave modes in connection with the existence of stationary waves (oscillitons). After the first description of oscillitons in a two-ion plasma [Sauer et al., 2001], other types of stationary waves were considered, for example in the range of whistler waves [Sauer et al., 2002], whose coupling with Langmuir waves [Sauer and Sydora, 2011] and EMIC waves in multi-ion plasmas [Sauer and Dubinin, 2022].

In the present case there are two possible types of stationary waves. At quasi-parallel propagation, as can be seen in Figure 5, panels $a_1$, $b_1$ and $d_1$, $e_1$ in the area of the dashed circles, mode splitting occurs as a result of the intersection of the L wave with the magneto-acoustic wave (MA). An additional condition is that the phase velocity of the MA wave remains below the Alfven velocity $V_{Ap}$ of the protons and that the damping does not exceed a certain value. Thus, the electromagnetic wave in the form of L-MA oscillitons can be expected to have the characteristics of both wave modes. Somewhat different conditions apply to the L-MA wave at quasi-perpendicular propagation, which is assigned to the observed density enhancements. As can be seen in the result of the Vlasov theory in Figure 5, panel $e_2$. the phase velocity of the wave has a maximum at $kc/\omega_p \sim 5$, which, similar to whistler waves [Sauer et al., 2002; Sydora et al., 2007; Dubinin et al., 2007] suggests the existence of stationary waves after the instability due to temperature anisotropy has subsided. It has to be pointed out that the oscilliton of the L-MA wave (identical with the kinetic Alfven wave) would be the only type so far considered which is of dominant electrostatic character.

The fact that the density enhancements are present in the vicinity of the large proton temperature anisotropy, but not exactly at the time of their appearance, is considered as a hint on the formation of stationary waves which survive after the saturation of the governing instability.

In summary of the theoretical analyses, an unstable magneto-acoustic mode is described that is consistent with the experimental requirements that the wave must propagate close to perpendicular to the background magnetic field in an environment having plasma betas close to 0.1 and a proton temperature anisotropy that is as large as eight. This theory produces an electromagnetic low frequency wave with a higher frequency electrostatic wave, just as observed in the described measurements.

In closing, it is worth noting that a similar wave structure exists in whistler mode wave events which consist of a low frequency electromagnetic wave and higher frequency whistler mode harmonics and spikes [Agapitov et al, 2018].

**V Acknowledgements**

This work was supported by NASA contract NNN06AA01C. The authors acknowledge the extraordinary contributions of the Parker Solar Probe spacecraft engineering team at the Applied Physics Laboratory at Johns Hopkins University. The FIELDS experiment on the Parker Solar Probe was designed and developed under NASA contract NNN06AA01C. Our sincere thanks to P. Harvey, K. Goetz, and M. Pulupa for managing the spacecraft commanding, data processing, and data analysis, which has become a heavy load thanks to the complexity of the instruments and the orbit. We also acknowledge the SWEAP team for providing the plasma density data. The work of I.V. was supported by NASA Heliophysics Guest Investigator grant 80NSSC21K0581

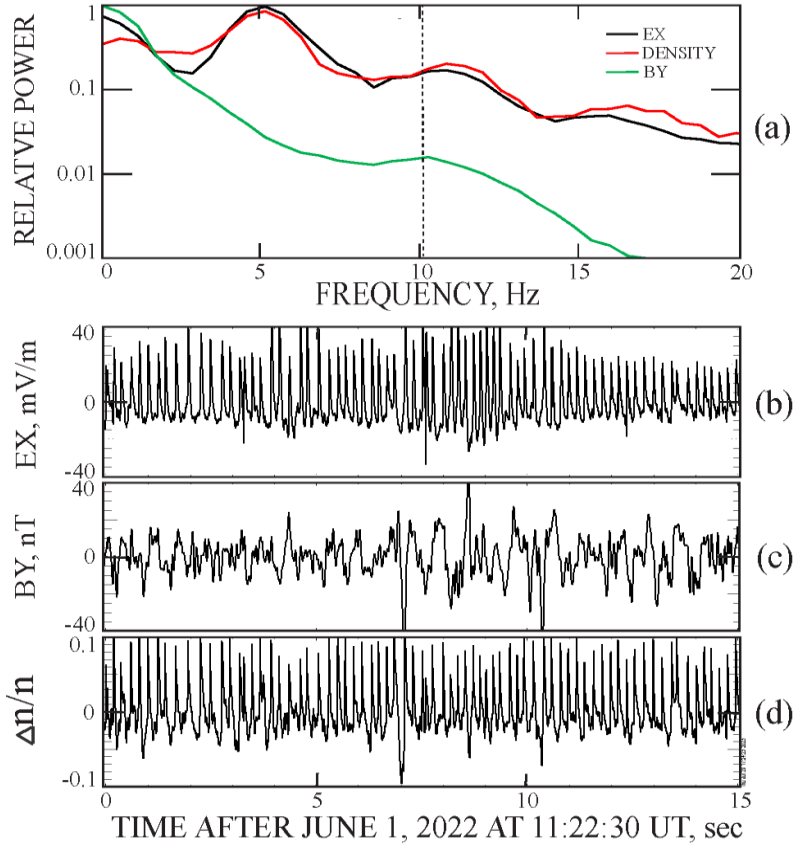

Figure 1. Power spectra of the electric field (black), plasma density (red), and magnetic field (green) in panel 1a at the time of the data in the remaining panels. The magnetic field spectrum appears normal with both a peak at ~1 Hz (normally attributed to Alfvenic turbulence), a large decrease of power between 1 and 20 Hz, and a break in the spectrum near 10 Hz, at a frequency near the proton gyrofrequency (the dashed vertical line). The electric field and density spectra in the panel 1a display peaks at ~1 Hz, 5 Hz and harmonics, and have a much smaller decrease with frequency than does the magnetic field. Because there is wave power in all three components near 1 Hz, this frequency signature must come from a mixture of electromagnetic and electrostatic waves, so it is not pure Alfvenic turbulence. The harmonic signatures at higher frequencies are purely electrostatic and the wave power is much greater than that expected for Alfvenic turbulence. The electric field, magnetic field, and density waveforms that produced these spectra are illustrated in panels 1b, 1c, and 1d.

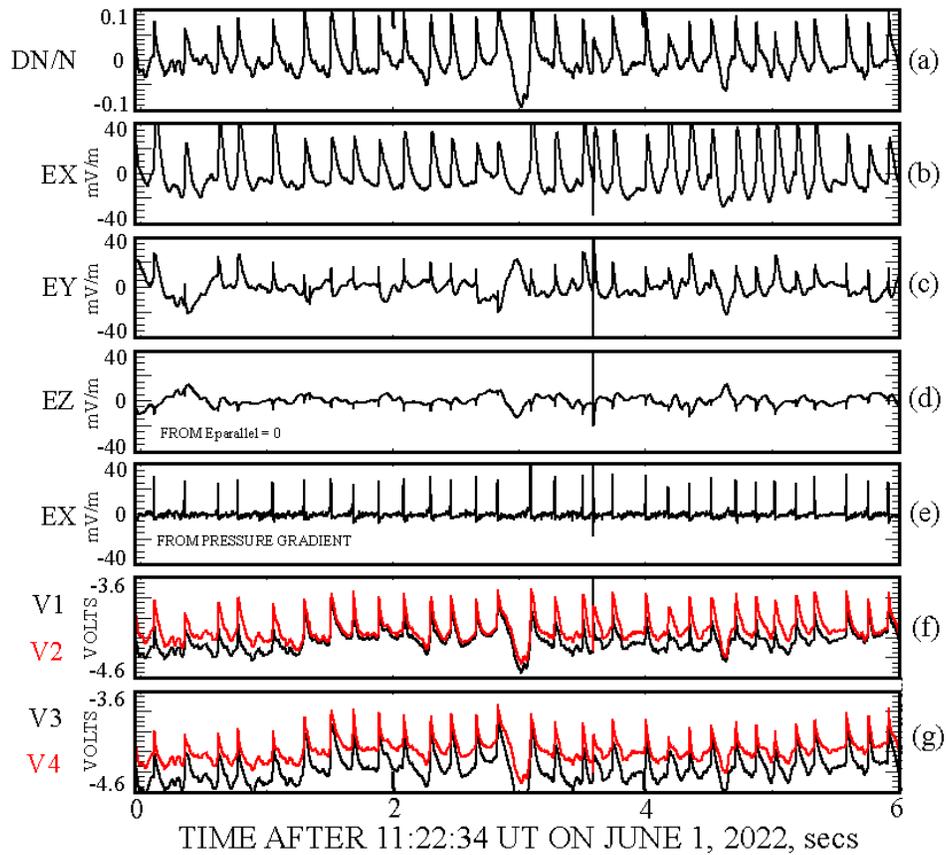

Figure 2. Density streams during a six second interval (panel 2a), measured X and Y components of the electric field (panels 2b and 2c), and an estimate of the unmeasured Ez (panel 2d), obtained from the two measured electric field components and the assumption that the parallel electric field was zero. The Ex electric field of panel 2e was computed from the Ohm's law requirement that the pressure gradient due to the density fluctuations of panel 2a was balanced by the electric field. It is noted the Ex was larger than the other two electric field components which shows that the electrostatic k-vector was in the X-direction, which was perpendicular to the background magnetic field. Also shown (panels 2f and 2g) are the four individual antenna voltages whose average is the spacecraft potential. All data in this figure, other than that in panels 2f and 2g, are high pass filtered at one Hz.

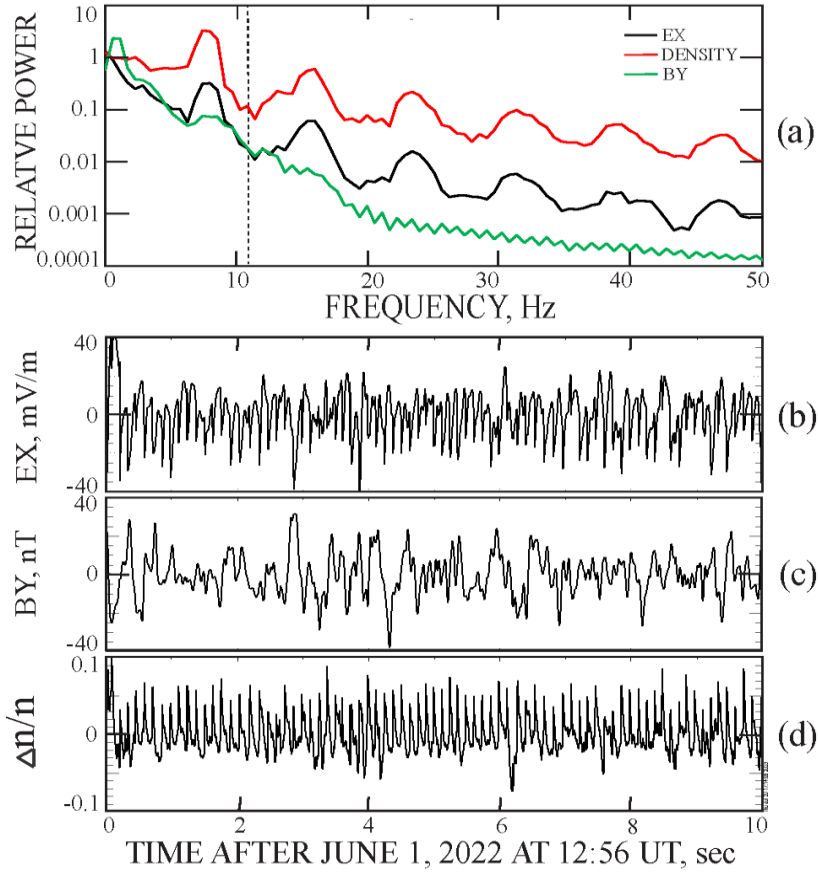

Figure 3. Ten seconds of data in panel 3a that illustrate the power spectra of the electric field (black), the density fluctuations (red), and the magnetic field (green). The one and seven Hz peaks contain electric field, magnetic field, and density maxima, indicating that both an electrostatic and electromagnetic wave existed at these frequencies. It is noted that the amplitude of the electrostatic wave at all frequencies was large compared to that of the Alfvenic turbulence expected from the magnetic field spectrum. Thus, this new wave mode may be an important contributor to the heating, scattering, and acceleration of the plasma.

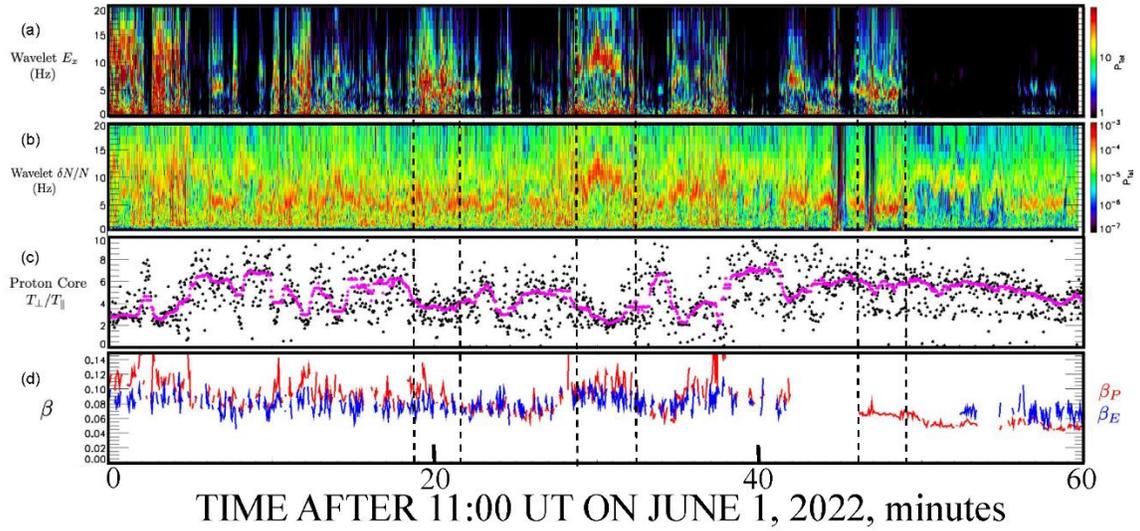

Figure 4. Panel 4a gives the electric field spectrum as a function of time, 4b gives the spectrum of density fluctuations, 4c gives the instantaneous perpendicular to parallel proton temperature ratio as the black dots and their 30-second-averages as the red dots, while 4d gives the electron and proton betas as computed from the quasi-thermal-noise. Note that the spike at 11:38 is an artifact. Three pairs of vertical dashed lines denote three of the about eight regions in which the proton and density fluctuations in panels 4a and 4b were large. These regions occurred near but not necessarily in regions where the temperature anisotropy in panel 4c was larger than six.
14

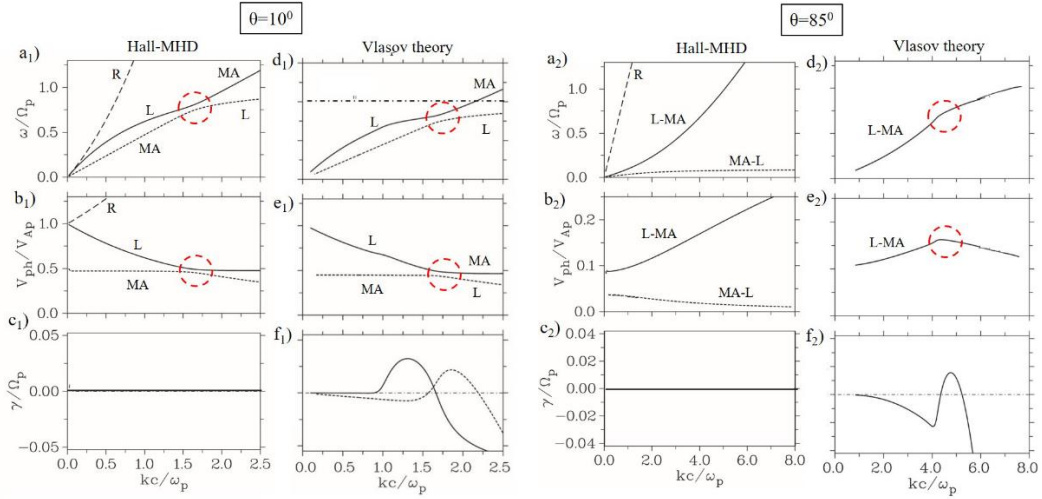

Figure 5. Dispersion of electromagnetic waves in the vicinity of the proton cyclotron frequency $\Omega_p$ based on the Hall-MHD and Vlasov theory for two propagation angles $\theta=10^0$ ($a_1$-$f_1$) and $\theta=85^0$ ($a_2$-$f_2$). From top to bottom, the normalized frequency $\omega/\Omega_p$, the phase velocity $V_{ph}$ normalized to the proton Alfven velocity $V_{Ap}$ and the growth rate $\gamma/\Omega_p$ versus the normalized wave number $kc/\omega_p$ are shown; $\omega_p$ is the proton plasma frequency. In the considered frequency band three wave modes occur. For quasi-parallel propagation these are the right-hand polarized mode (R), the left-hand-polarized ion cyclotron mode (L) and the magneto-acoustic mode (MA). The mixed character of the L and MA modes for quasi-perpendicular propagation is expressed by the designations L-MA (kinetic Alfven wave) and MA-L. The following parameters have been used: $\beta_e=0.15$, $\beta_p=0$ in Hall-MHD and $\beta_e=0.15$, $\beta_p=0.15$ in Vlasov theory; $\beta_e$ and $\beta_p$ are the electron plasma beta of electrons and protons, respectively. The temperature anisotropy which drives the kinetic instabilities is for both cases ($\theta=10^0$ and $\theta=85^0$) the same, namely $T_{perp}/T_{par}=6$.